%Paper: solv-int/9510004
%From: Richard Ward <Richard.Ward@durham.ac.uk>
%Date: Tue, 17 Oct 1995 17:06:07 +0100 (BST)

\magnification=\magstep1
\baselineskip=20 truept
\def\ni{\noindent}

\def\bn{\bigskip\noindent}
\def\sn{\smallskip\noindent}
\def\pa{\partial}
\def\ee{{\rm e}}  %exponential
\def\fs{\vert f\vert^2}

\def\ve{\varepsilon}

\font\scap=cmcsc10
\font\title=cmbx10 scaled\magstep1

\def\ps{\Vert p\Vert^2}
\def\qs{\Vert q\Vert^2}
\def\zb{\bar z}
\def\JI{J^{-1}}

\rightline{DTP95/59}
\vskip 1truein
\centerline{\title Nontrivial scattering of localized solitons}
\centerline{\title in a (2+1)-dimensional integrable system.}

\vskip 1truein
\centerline{\scap R. S. Ward}

\bn\centerline{\sl Dept of Mathematical Sciences, University of Durham,}
\centerline{\sl Durham DH1 3LE, UK.}

\vskip 2truein
\ni{\bf Abstract.} One usually expects localized solitons in an integrable
system to interact trivially. There is an integrable (2+1)-dimensional
chiral equation which admits multi-soliton solutions with trivial dynamics.
This paper describes how to generate explicit solutions representing nontrivial
soliton interactions: in particular, a head-on collision of two solitons
resulting in $90^\circ$ scattering.

\bn{\it To appear in Physics Letters A.}

\vfill\eject

\ni{\bf1. Introduction.}

\sn This paper is concerned with localized solitons in the plane,
{\it i.e.\/}~in three-dimensional space-time. Systems admitting such solitons
may be grouped into various distinct classes.
For example, there are systems admitting
topological solitons, the stability of which depends on nontrivial
topology: these include vortices in the abelian Higgs model [1], and lump
solutions of sigma-models (with various possible modifications) [2].
These are not solitons in the strictest sense; for example, the collision of
two solitons is not elastic (some radiation is emitted). The feature of
topological systems which is relevant here, is that a head-on collision
results in $90^\circ$ scattering. In other words, if two solitons approach
each other along the $x$-axis and collide at the origin, then two solitons
emerge, travelling in opposite directions along the $y$-axis (with slightly
less speed, because of the inelasticity). When they overlap at the origin,
they form a ring rather than a single lump. This phenomenon has been observed
in numerical simulations [3--7] and understood analytically [8--13].

Some other systems that admit localized solitons are those that are
completely integrable. One long-standing example is the
Kadomtsev-Petviashvili equation. In this case, there are solutions
representing $n$ solitons, the interaction of which is completely trivial
[14]. In particular, if two solitons collide head-on, then they emerge
in the same direction (without even a phase shift), and with the same speeds.
Another system in which exactly the same thing happens is an integrable
chiral model [15], which is the subject of this paper, and which will be
described below (see eqn 1). This system is closely related to one
introduced in [16], where once again (to quote from that paper)
``two-dimensional solitons do not
interact''. One might be tempted to conclude that for integrable systems,
localized solitons on the plane have trivial dynamics. (There are
exceptions such as dromion solutions of the Davey-Stewartson equations,
but these are driven by nontrivial boundary conditions at spatial infinity.)

This picture was, however, undermined by some numerical solutions of the
integrable chiral equation [17, 7]. These revealed that solitons can also
scatter at right angles, as in the topological models referred to previously.
Solitons in this system have internal degrees of freedom, and the numerical
results seemed to indicate that solitons can interact either trivially or
nontrivially, depending on the orientation of these internal parameters.
Since the system is integrable, one might expect there to be explicit
solutions which exhibit nontrivial scattering. This paper provides an example
of such a solution, and gives a systematic way of constructing many more.

We must begin, however, by setting up the system. It is defined on
(2+1)-dimensional space-time, with coordinates $(t,x,y)$ where $t$ denotes
time. The chiral field $J(t,x,y)$ is a $2\times2$ unitary matrix with
$\det J=1$, and its equation of motion is
$$
\pa_t(\JI J_t) - \pa_x(\JI J_x) - \pa_y(\JI J_y)
   + [\JI J_y, \JI J_t] = 0. \eqno(1)
$$
Here $\pa_t=\pa/\pa t$, $J_t=\pa J/\pa t$, etc. The standard SU(2) chiral
equation has only the first three terms of (1), and not the commutator term:
but the latter equation is not integrable. By contrast, eqn (1) is a
dimensional reduction of the self-dual Yang-Mills equations in 2+2 dimensions;
it is associated with a linear system (see eqn 6), has an inverse scattering
description [18], passes the Painlev\'e test for integrability [19],
and so forth.

The energy of $J$ is defined to be $E=\int\!\!\!\int{\cal E}\,dx\,dy$, where
the energy density ${\cal E}$ is
$$
{\cal E} = -{\rm tr}\bigl[(\JI J_t)^2+(\JI J_x)^2+(\JI J_y)^2\bigr].
$$
This quantity $E$ is a positive-definite functional of $J$, and is conserved
[15]. In order to ensure finite energy, we impose on $J$
the boundary condition
$$
J = J_0 + J_1(\theta)r^{-1} + O(r^{-2})\quad{\rm as}\ r\to\infty, \eqno(2)
$$
where $x+iy=r\exp(i\theta)$. The matrix $J_0$ is constant, and $J_1$ is allowed
to depend only on~$\theta$ (not on~$t$).

The simplest nontrivial solution of this system is the static 1-soliton
located at the origin, which is given by
$$
  J = i \Biggl(I-{2p^*\otimes p \over\ps}\Biggr).
$$
Here $I$ is the identity $2\times2$ matrix, $p$ is the 2-dimensional
row vector $(1,z)$ where $z=x+iy$, $p^*$ denotes the complex conjugate
transpose of $p$, and $\ps=p\cdot p^*$ is the norm-squared of $p$.
There is numerical evidence that this solution is stable [20].
The energy density goes like ${\cal E}=O(r^{-4})$ as $r\to\infty$, which is
the case for all the solutions described in this paper.

\vfill\eject

\ni{\bf2. A Scattering Solution.}

\sn In this section, we shall see an example of a solution of (1) representing
two solitons which undergo $90^\circ$ scattering.
The solution has the form of a product
$$
  J = \Biggl(I-{2p^*\otimes p \over\ps}\Biggr)
      \Biggl(I-{2q^*\otimes q \over\qs}\Biggr), \eqno(3)
$$
where $p$ and $q$ are the row vectors
$$\eqalign{
           p &= (1,z), \cr
           q &= (1+r^2)\,(1,z) - 2ig\,(\zb, 1), \cr} \eqno(4)
$$
with $g=t+z^2$. This field $J$ takes values in SU(2); is smooth everywhere
on space-time (note that $p$ and $q$ are nowhere-zero, which is necessary
for smoothness); satisfies the boundary condition (2); and satisfies the
equation of motion (1) (this follows from the construction in the next
section).

The crucial features of this time-dependent solution may be inferred as
follows. If $r^2=z\zb$ is large, then $J$ is close to its asymptotic
value $J_0$, as long as $g/z^2 = 1+t/z^2 \approx 1$. But as $z$ approaches
$\pm\sqrt{-t}$, {\it i.e.\/}~as $g\to0$, $J$ departs from its asymptotic value:
this is approximately where the solitons are located. If, therefore, $t$ is
negative, then there are two solitons, on the $x$-axis at
$x\approx\pm\sqrt{-t}$; while if $t$ is positive, then there are two solitons
on the $y$-axis, at $y\approx\pm\sqrt{t}$.
So the picture is of two solitons accelerating towards each other, scattering
at right angles, and then decelerating as they separate. They accelerate
as if there were a mutual attractive force, proportional to the
inverse cube of the distance between them.

This picture can be confirmed by looking in more detail at the energy density
of the solution, which is
$$
  {\cal E} = 32{1+10r^2+5r^4+4t^2(1+2r^2)-8t(x^2-y^2)
                \over \bigl[1+2r^2+5r^4+4t^2+8t(x^2-y^2)\bigr]^2}.\eqno(5)
$$
Note, first, the symmetry of ${\cal E}$ under the interchange $t\mapsto -t$,
$x\mapsto y$, $y\mapsto x$: the collision is time-symmetric
and elastic in this sense,
and produces no radiation. For large (positive) $t$, ${\cal E}$ is peaked,
as expected, at two points on the $y$-axis, namely $y\approx\pm0.92\sqrt{t}$.
The corresponding solitons are not, however, of constant size: their
height (the maximum value of ${\cal E})$ is proportional to $1/t$,
while their radii are proportional to $\sqrt{t}$. (The total ``volume'',
{\it i.e.\/}~the spatial integral of ${\cal E}$, is of course constant in
time.) In other words, the solitons spread out as they move apart.

Figure 1 illustrates what happens near $t=0$. The solitons coming in along
the $y$-axis merge to form a ring, and then emerge along the $x$-axis.

\bn{\bf3. Construction of Solutions.}
\sn This section will indicate how the solution (3) was constructed, and
how other solutions may be obtained. The procedure is a variant of that in
[15]; the latter method was in turn pioneered by Zakharov {\it et al\/}
[21, 22].

The nonlinear equation (1) is the consistency condition for a pair of linear
equations for a $2\times2$ matrix $\psi(\zeta)$, where $\zeta$ is a complex
variable ($\psi$ also depends on the space-time variables $t$, $x$, $y$).
This linear pair is
$$\eqalign{
   D_1\psi := (\zeta\pa_x-\pa_y-\pa_t)\psi     &= A_1\psi, \cr
   D_2\psi :=(\zeta\pa_t-\zeta\pa_y-\pa_x)\psi &= A_2\psi, \cr
}\eqno(6)$$
where $A_1$ and $A_2$ are $2\times2$ matrices which are independent of $\zeta$.
Indeed, the integrability condition for (6) implies that there exists a $J$
such that
$$
  A_1=\JI(J_t+J_y), \qquad A_2=\JI J_x; \eqno(7)
$$
and that this $J$ satisfies the equation of motion (1). Comparing (6) and
(7), we see that $J$ can be identified with $\psi(0)^{-1}$.

The unitarity condition on $J$ follows from an analogous condition on
$\psi(\zeta)$, namely
$$
  \psi(\zeta) \psi(\bar\zeta)^* = I. \eqno(8)
$$
So the idea is that if we can find a $\psi(\zeta)$ such that the unitarity
condition (8) holds, and such that $(D_i\psi)\psi^{-1}$ is independent of
$\zeta$ for $i=1,2$, then $J=\psi(0)^{-1}$ is a unitary solution of (1).

The standard way of constructing multi-soliton solutions is to assume that
$\psi$ has simple poles in $\zeta$, in fact that $\psi$ has the form
$$
  \psi(\zeta) = I + \sum_{k=1}^n {M_k \over \zeta-\mu_k}, \eqno(9)
$$
where the $M_k$ are matrices independent of $\zeta$. This leads to an
$n$-soliton solution, in which the velocity of the $k$th soliton is
determined by the constant complex number $\mu_k$; one consequence is that
there is no scattering [15, 16].

The scattering solution (3), by contrast, arises from a $\psi$ with a double
pole (and no other poles) in $\zeta$. In other words, $\psi$ is taken to have
the form
$$
\psi(\zeta) = I + {R_1\over\zeta-\mu} + {R_2\over(\zeta-\mu)^2}. \eqno(10)
$$
The first requirement on this $\psi$ is the unitarity condition (8). By
examining the coefficients of the (apparent) poles of
$\psi(\zeta) \psi(\bar\zeta)^*$, one may easily show that $\psi$ satisfies (8)
if and only if it factorizes as
$$
\psi(\zeta) = \Biggl(I-{(\bar\mu-\mu)\over(\zeta-\mu)}{q^*\otimes q\over\qs}
                \Biggr)\Biggl(I-{(\bar\mu-\mu)\over(\zeta-\mu)}
              {p^*\otimes p \over\ps}\Biggr) \eqno(11)
$$
for some 2-vectors $p$ and $q$. In fact, the same is true if $\psi(\zeta)$
has a pole of order $n$: the unitarity condition is satisfied if and only if
$\psi$ factorizes into $n$ simple factors of the type appearing in (11)
(proof by induction on $n$).

Next, we have to ensure that $(D_i\psi)\psi^{-1}$ is independent of $\zeta$:
this imposes differential equations on $p$ and $q$. These are coupled
nonlinear equations, and it seems difficult to find their general solution.
One way of proceeding is to take a solution for the simple-pole case (9) with
$n=2$, of which many are known, and to let $\mu_k\to\mu$ for $k=1,2$.
If we arrange things carefully, this limit gives a solution of the double-pole
type (11). (It may seem strange that one can take the limit of a family
of two-soliton solutions with trivial scattering, and obtain a two-soliton
solution with highly non-trivial scattering; but in fact this is exactly
what happens.)

For the $n$-soliton solution corresponding to (9), each matrix $M_k$ is given
in terms of a rational meromorphic function $f_k$ of one complex variable [15].
(Roughly speaking, $f_k$ describes the shape of the $k$th soliton.) In our
case, with $n=2$, we put $\mu_1=\mu+\ve$, $\mu_2=\mu-\ve$ and take the limit
$\ve\to0$. In order for the resulting $\psi$ to be smooth for all $(t,x,y)$,
it is necessary that $f_2-f_1\to0$ as $\ve\to0$. So let us write
$f_1=f+\ve h$, $f_2=f-\ve h$, where $f$ and $h$ are rational functions of
one variable. On taking the limit $\ve\to0$, we then obtain a $\psi$ of the
form (11), smooth for all $(t,x,y)$, and satisfying the requirement that
$(D_i\psi)\psi^{-1}$ be independent of $\zeta$. Consequently, $J=\psi(0)^{-1}$
is a smooth unitary solution of~(1).

So we have a solution depending on the parameter $\mu$ and on the two
arbitrary rational functions $f$ and $h$. To get the solution of the previous
section, take $\mu=i$ (which corresponds to the ``centre-of-mass''
of the system being stationary); this yields (3), with $p$ and $q$ being given
in terms of $f(z)$ and $h(z)$ by
$$\eqalign{
   p &= (1,f), \cr
   q &= (1+\fs)\,(1,f) -2ig\,(\bar f, -1), \cr} \eqno(12)
$$
where $g=tf'(z)+h(z)$. To get the particular form (4) used in the previous
section, one takes
$$
  f(z) = z, \qquad h(z) = z^2. \eqno(13)
$$

So the scattering solution belongs to a large family: one may take $f$ and $h$
to be any rational meromorphic functions of $z$. It is not difficult to
check that $J$ satisfies its boundary condition as $r\to\infty$, irrespective
of the choice of $f$ and $h$.

The equation (1) does not have rotational symmetry in the $xy$-plane. So one
might suspect that the solution (3,4) is somehow exceptional, involving as it
does solitons moving along the coordinate axes. But in fact one can rotate
the whole picture of figure 1 through any angle $\phi$
in the $xy$-plane, by making the choice
$$
f(z)=z, \qquad h(z)=\ee^{2i\phi}z^2. \eqno(14)
$$
The energy density of the solution corresponding to (14) is the same as in (5),
except that $x^2-y^2=\Re[z^2]$ is replaced by $\Re[\exp(2i\phi)z^2]$.
Even though the equation is not rotationally symmetric, one can
find a new solution which is a rotated version of the original solution.
A possible way of understanding this
is to note that (1) can be written in a Lorentz-invariant form, as a
Yang-Mills-Higgs equation [23, 20]; so it has a ``hidden'' Lorentz symmetry.

\bn{\bf4. Concluding Remarks.}

\sn The solution described above, with $\psi(\zeta)$ having a double pole,
is an example of a 2-uniton. The idea of $n$-unitons was introduced in
connection with finding SU($N$) chiral fields on R$^2$ [24], and it extends
naturally to the corresponding system on R$^{2+1}$ [23]. For SU(2) chiral
fields on R$^2$, {\it i.e.\/}~the static version of the system of this paper,
one uniton is enough (the static soliton is a 1-uniton). The original
reason for introducing $n$-unitons was that they are needed for static
SU($n+1$) chiral fields [24]. What we have seen is that higher unitons are
also needed for the time-dependent SU(2) case.

It seems likely that there are many more interesting solutions still to be
revealed, both in the 2-uniton class discussed in the previous section,
and corresponding to higher unitons ($\psi$ having a higher-order pole in
$\zeta$). One could, for example, ask what happens when the impact
parameter of the collision is nonzero; and whether the sizes of interacting
solitons must necessarily be nonconstant. In addition, it would be interesting
to elucidate the role of higher unitons in the inverse-scattering [16,18]
and algebraic-geometry [23] approaches to this system.

\vfill\eject

\ni{\bf References.}
\smallskip
\item{[1]} A. Jaffe and C. Taubes, Vortices and monopoles (Birkh\"auser,
       Boston, 1980).
\item{[2]} R. S. Ward, in: Harmonic maps and integrable systems, eds A. P.
           Fordy and J. C. Wood (Vieweg, Wiesbaden, 1994) p. 193.
\item{[3]} K. J. M. Moriarty, E. Myers and C. Rebbi, Phys. Lett. B 207
           (1988) 411.
\item{[4]} E. P. S. Shellard and P. J. Ruback, Phys. Lett. B 209 (1988) 262.
\item{[5]} R. A. Leese, M. Peyrard and W. J. Zakrzewski, Nonlinearity 3
           (1990) 773.
\item{[6]} W. J. Zakrzewski, Nonlinearity 4 (1991) 429.
\item{[7]} B. Piette, P. M. Sutcliffe and W. J. Zakrzewski, Int. J. Mod.
           Phys. C 3 (1992) 637.
\item{[8]} P. J. Ruback, Nucl. Phys. B 296 (1988) 669.
\item{[9]} R. Leese, Nucl. Phys. B 344 (1990) 33.
\item{[10]} P. M. Sutcliffe, Nonlinearity 4 (1991) 1109.
\item{[11]} C. Rosenzweig and A. M. Srivastava, Phys. Rev. D 43 (1991) 4029.
\item{[12]} I. A. B. Strachan, J. Math. Phys. 33 (1992) 102.
\item{[13]} A. Kudryavtsev, B. Piette and W. J. Zakrzewski, Phys. Lett. A
            180 (1993) 119.
\item{[14]} S. V. Manakov, V. E. Zakharov, L. A. Bordag, A. R. Its and
            V. B. Matveev, Phys. Lett. A 63 (1977) 205.
\item{[15]} R. S. Ward, J. Math. Phys. 29 (1988) 386.
\item{[16]} S. V. Manakov and V. E. Zakharov, Lett. Math. Phys. 5 (1981) 247.
\item{[17]} P. M. Sutcliffe, J. Math. Phys. 33 (1992) 2269.
\item{[18]} J. Villaroel, Stud. Appl. Math. 83 (1990) 211.
\item{[19]} R. S. Ward, Nonlinearity 1 (1988) 671.
\item{[20]} P. M. Sutcliffe, Phys. Rev. D 47 (1993) 5470.
\item{[21]} A. A. Belavin and V. E. Zakharov, Phys. Lett. B 73 (1978) 53.
\item{[22]} V. E. Zakharov and A. B. Shabat, Func. Anal. Appl. 13 (1980) 166.
\item{[23]} R. S. Ward, Commun. Math. Phys. 128 (1990) 319.
\item{[24]} K. K. Uhlenbeck, J. Diff. Geom. 30 (1989) 1.

\bye